\documentclass{ws-procs9x6}

\begin{document}

\title{BL Lac X-ray Spectra:  simpler than we thought}

\author{Eric S. Perlman, Timothy Daugherty, Anuradha Koratkar}

\address{Joint Ctr for Astrophysics,
University of Maryland--Baltimore County}

\author{Grzegorz Madejski, Karl Andersson}

\address{ Stanford University, SLAC}

\author{Julian H. Krolik}

\address{Dept. of Physics and Astronomy, Johns Hopkins University}

\author{Hugh Aller, Margo Aller}

\address{Department of Astronomy, University of Michigan}

\author{John T. Stocke}

\address{Center for Astrophysics and Space Astronomy,
University of Colorado}

\author{Travis Rector}

\address{Department of Physics and Astronomy,
University of Alaska -- Anchorage}

\author{Paolo Padovani}

\address{European Southern Observatory}

\author{Alan Marscher}

\address{Department of Astronomy, Boston University}

\author{Mark Allen}

\address{Centre de Donnees Astronomique}

\author{Stefan Wagner}

\address{Landessternwarte Heidelberg}


\maketitle

\abstracts{ We report results from {\it XMM-Newton} observations of thirteen
X-ray bright BL Lacertae objects, selected from the {\it Einstein} Slew Survey
sample.   The spectra are generally well fit by power-law models, with four
objects having hard ($\alpha<1; F_\nu \propto \nu^{-\alpha}$) spectra that
indicates synchrotron peaks at $>5$ keV.  None of our spectra show line
features, indicating that soft X-ray absorption ``notches'' must be rare
amongst BL Lacs, rather than common or ubiquitous as had previously been
asserted.  We find significant curvature in most of the spectra.  This
curvature is almost certainly intrinsic, as it appears nearly constant 
from  0.5 to 6 keV, an observation which is inconsistent with the small
columns seen in these sources.}

\section{Introduction}

The nature of the X-ray emission and absorption from BL Lacs is still an open
question.  Most BL Lac objects have X-ray spectra that can be fit by power-laws
within  smaller bandpasses, such as that of {\it ROSAT} ([12],[17]) More recent
results from {\it ASCA} [7] and {\it BeppoSAX} ([1],[11],[19]) have generally
confirmed this spectral morphology, and also added onto it the possibility of
intrinisic spectral curvature across a wider bandpass ([4],[6],[9],[16]). In
addition, earlier missions had indicated that some BL Lac objects showed a
deficit in soft X-rays below a power-law model, which had been interpreted by
invoking X-ray absorption features at 0.5--0.8 keV ([3],[8],[14],[15]).  But
not all bright BL Lacs were found to require such features  (e.g., [5]).  

\section{Sample, Observations and Data Reduction}

We selected our targets from the {\it Einstein} Slew Survey sample of BL Lacs
[13].  The Slew Survey sample is the largest collection  of
X-ray bright BL Lacs, and was the first contain  significant numbers of both
HBLs and LBLs (high-energy and low-energy peaked BL Lacs; Urry \& Padovani
1995).   We received  time for the 13 X-ray brightest objects (all HBLs) 
which were not on either {\it XMM} GTO lists or in the  Cycle 1 {\it
Chandra} schedule.   Seventeen observations were done (four were repeated
due to background flares). Because the main goal was to address the class 
properties, our integration times were short, $\sim 5$ ks.
As a result, the best data come from the EPIC instruments, which are imaging
spectrographs that have low-to-moderate ($R\sim20-50$) spectral resolution.

All source and background extraction was done in SAS v5.4.1.   X-ray spectral
modeling was done in XSPEC v11.0.  The PN data were fit between 1.1-10.0 keV, 
while the MOS data were fit between 0.5-10.0 keV, except for the faintest
objects which we capped at 7.0 keV due to low count rates at the highest
energies. Where multiple observations of an object were obtained, each
observation was reduced and analyzed separately. Three models were fit:  a
single power law, sum of two power laws, and a logarithmic parabola. Each  
model was attempted with  Galactic and variable absorption, and we also
fitted several sub-bands to investigate curvature.

\section{Results}

The spectral indices we found range from $\Gamma=1.7-3.5$, with 12/17 being in
the range $\Gamma=2.0-2.9$.  This is similar to previous findings.  Four
spectra were found to be flat ($\Gamma<2$).  These objects are likely to have
$\nu_{peak}> 5$ keV. In 14/17 observations, a better fit was obtained by
allowing for  spectral curvature, which may be  intrinsic, or the result of
additional absorption.  We believe the curvature is most likely intrinsic,
because it appears nearly constant between 0.5-6 keV in all objects where it is
required.  This is inconsistent with absorption given the observed columns,
which range from  $3-20 \times 10^{20} {\rm ~cm^{-2}}$.  The curvature can be
characterized by  $d\Gamma/d(\log E)\approx 0.4 \pm 0.15$.  

A similar curvature was found by Giommi et al. [4] for about 50\% of X-ray
bright BL Lacs observed by {\it BeppoSAX}. The greater percentage we find to
require curvature, is consistent with the greater sensitivity of XMM. 
This type of curvature has more recently been analyzed ([9],[10]) in the 
context of particle acceleration models, and given the multiple emission
regions that most likely contribute to BL Lac X-ray spectra, a continuous
curvature is the most likely result of spectral aging in several regions with
different physical  parameters (as opposed to the simpler model of a sharp
cutoff, which is more commonly assumed).

The finding of significant absorption features, stands in stark contrast to the
claims of "ubiquitous absorption features" in the spectra of BL Lac objects,
made by earlier workers. We are confident of this result based on the high
signal-to-noise of these spectra.  In addition, another, independent study of
four of the five BL Lacs where BBXRT and ASCA spectra appeared to give these
line features[2], found a similar lack
of features. We believe the most consistent explanation for the earlier results
is that those spectra did indeed show curvature, similar to what we see in our
spectra, but interpreted that curvature incorrectly as absorption.

\begin{figure}[ht]

\centerline{\epsfxsize=3.7in\epsfbox{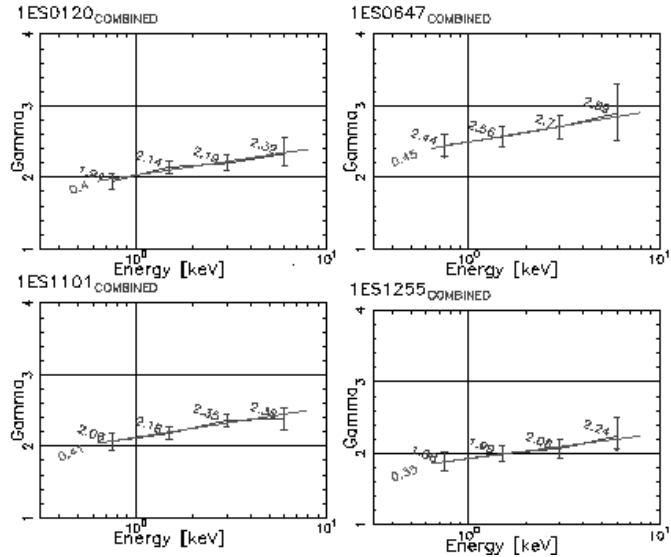}}
\caption{Four examples of the curved X-ray spectra seen for the objects
in our sample.  All of these objects show steeper spectra at higher energies, 
with curvature that remains constant through at least 6 keV.}

\end{figure}

\end{document}